\documentclass{blois}
\usepackage{microtype}

\def\be{\begin{equation}}
\def\ee{\end{equation}}
\def\bea{\begin{eqnarray}}
\def\eea{\end{eqnarray}}

\newcommand{\bbonu}{\ensuremath{0\nu\beta\beta}}

\newcommand{\Photo}{}

\begin{document}
\vspace*{4cm}

\title{THE NEXT EXPERIMENT FOR\\NEUTRINOLESS DOUBLE BETA DECAY SEARCHES}

\author{\sc Justo Mart\'in-Albo\thinspace\footnote{Now at the Instituto de F\'isica Corpuscular (IFIC), Universitat de Val\`encia \& CSIC, Valencia, Spain.} \\for the NEXT Collaboration}

\address{Department of Physics, Harvard University \\
17 Oxford St, Cambridge, MA 02138, USA}

\maketitle

\abstracts{
The \emph{Neutrino Experiment with a Xenon TPC} (NEXT) seeks to discover the neutrinoless double beta (\bbonu) decay of $^{136}$Xe using a high-pressure xenon gas time projection chamber with electroluminescent amplification. The observation of \bbonu\ decay would prove that neutrinos are Majorana particles and that lepton number is not conserved. The NEXT detector concept offers several features of great value for \bbonu-decay searches, including excellent energy resolution, tracking for the active suppression of backgrounds and scalability to large source masses. The initial phase of the NEXT project (2009--2014) was devoted to R\&D with two prototypes (DEMO and DBDM) of approximately 1~kg of active xenon mass that demonstrated the performance of the detector concept. During the second phase of the project (2015--2019), the NEXT Collaboration has operated underground at the \emph{Laboratorio Subterr\'aneo de Canfranc}, in Spain, a radio-pure detector of about 5~kg of xenon mass. The goal of the current phase is the construction, commissioning and operation of the NEXT-100 detector, with a predicted sensitivity to the \bbonu\ decay half-life of $6\times10^{25}$~years (90\% CL) after a run of 3 years. The Collaboration is planning as well a future tonne-scale phase to explore $\bbonu$-decay half-lives beyond $10^{26}$~years.
}

\section{Introduction} \label{sec:Introduction}
Neutrinoless double beta (\bbonu) decay is a hypothetical second-order weak process in which a nucleus of atomic number $Z$ and mass number $A$ transforms into its $Z+2$ isobar emitting two electrons. The discovery of this process would prove that neutrinos are Majorana particles ---\thinspace truly neutral particles identical to their antiparticles\thinspace--- and that total lepton number is not conserved in nature, two findings with far-reaching implications in particle physics and cosmology. Majorana neutrinos, for example, imply the existence of a new energy scale at a level inversely proportional to the observed neutrino masses.\cite{Weinberg:1979sa} Such a scale provides a simple explanation for the striking lightness of neutrinos,\cite{seesaw} 
and is probably connected to several other open questions in particle physics. Moreover, Majorana neutrinos could have induced the cosmological asymmetry between matter and antimatter through a mechanism knows as \emph{leptogenesis}.\cite{Fukugita:1986hr}

Experimentally, no evidence of the existence of \bbonu\ decay has been obtained so far.\cite{GomezCadenas:2011it} The best limit to the half-life of the transition in $^{136}$Xe was set by the KamLAND-Zen experiment with an exposure of 504~kg~yr: $T^{0\nu}_{1/2} > 1.07\times10^{26}$~years at 90\% CL.\cite{KamLAND-Zen:2016pfg} The experimental goal for the coming decade is the exploration of the region of half-lives up to $10^{28}$~years, which will require exposures well beyond $1$~tonne~yr and background rates of the order of 1~count~tonne$^{-1}$~yr$^{-1}$.

\section{The NEXT project so far} \label{sec:TheNextProjectSoFar}
The \emph{Neutrino Experiment with a Xenon TPC} (NEXT) seeks to discover the \bbonu\ decay of $^{136}$Xe using a high-pressure xenon gas (HPXe) time projection chamber (TPC) with electroluminescent amplification. The experiment is a collaboration involving about 85 scientists from 20 different research institutes and universities in Spain, USA, Portugal and Colombia. The NEXT detector concept offers two features of great value for \bbonu-decay searches: good energy resolution ($<1\%$ FWHM at the energy of the transition, 2.5~MeV) and charged-particle tracking for the active suppression of background. Furthermore, the technology can be extrapolated to large source masses, allowing the exploration of the region of half-lives beyond 10$^{26}$~years. 

The detection principle of NEXT is as follows. Charged particles interacting in the HPXe transfer their energy to the medium through ionization and excitation. The excitation energy is manifested in the prompt emission of ultraviolet ($\sim$175 nm) scintillation light. The ionization tracks (positive ions and free electrons) left behind by the particle are prevented from recombination by a strong electric field. Negative charge carriers drift then toward the TPC anode, entering a region with an even more intense electric field. There, further ultraviolet photons are generated isotropically by the electroluminescence (EL) process. Therefore, both scintillation and ionization produce an optical signal, to be detected with photosensors arrays (PMTs or SiPMs) located behind cathode and anode. The detection of the primary scintillation light signals the start-of-event time, whereas the EL light is used for calorimetry and track reconstruction. 

The first phase of the NEXT project (2009--2014) was devoted to R\&D with two prototypes, NEXT-DBDM and NEXT-DEMO, of about 1~kg of active xenon mass. With them, the Collaboration assessed the detector concept and gained technical expertise that facilitated the design, construction and operation of larger systems.\cite{DBDM,DEMO}

The second phase of the project (2015--2019) has involved the underground operation at the \emph{Laboratorio Subterr\'aneo de Canfranc} (LSC), in the Spanish Pyrenees, of a 5-kg radio-pure detector called NEXT-{\sc White}.\cite{Monrabal:2018xlr} Its active volume is a cylinder 530~mm long and 454 mm in diameter. Two arrays of photosensors are located at opposite ends of the cylinder: the energy plane, composed of 12 low-background, 3-inch photomultiplier tubes (Hamamatsu R11410-10) is positioned behind the cathode, whereas the tracking plane, consisting of 1-mm$^2$ SensL silicon photomultipliers (SiPMs) spaced at a pitch of 1 cm, is located close behind the anode. The active volume is shielded from external radiation by a thick copper shield. This inner shield, the electric-field cage and both sensor planes are housed in a stainless steel pressure vessel designed to withstand up to 20 bar. The vessel sits on top of an anti-seismic pedestal and inside a lead shield.

The operation of NEXT-{\sc White} has established a procedure to calibrate the detector using $^{83m}$Kr decays, and provided initial measurements of energy resolution, electron drift parameters and a measurement of the impact of $^{222}$Rn in the radioactive budget.\cite{NEXT-White:2018} The most recent results of this detector include the measurement of an energy resolution at 2.5~MeV better than 1\% FWHM, the demonstration from the data themselves of a robust discrimination between 2-electron tracks (such as a double beta decay) and single electrons (the main background events), and a measurement of the radiogenic backgrounds, which demonstrates both the low radioactive budget of the apparatus and the accuracy of the background model.\cite{NEXT-White:2019}

\section{The future of the NEXT program} \label{sec:TheFuture}
The NEXT-100 detector, scheduled to start data taking in 2020, is the third phase of the program. It is a radio-pure HPXe TPC containing, at 15 bar pressure, about 100~kg of xenon enriched at 91\% in $^{136}$Xe. The active region of the detector is a cylinder of 1050~mm diameter and 1300~mm length (1.27~m$^3$ of active volume). The energy of the event is measured with an array of 60 Hamamatsu R11410-10 photomultiplier tubes. In addition, these PMTs record the primary scintillation that signals the initial time of the event. EL light is also detected a few mm away from production at the anode plane by a matrix of $\sim$5600 SiPMs. The combination of good energy resolution, tracking-based signal discrimination and low radioactive budget results in a very low expected background index of at most $4\times10^{-4}$ counts keV$^{-1}$ kg$^{-1}$ yr$^{-1}$. This translates into a sensitivity to the \bbonu\ decay half-life of $6\times10^{25}$~years (90\% CL) after a run of 3 years.\cite{Martin-Albo:2015rhw}

The NEXT technology can be scaled up to tonne-scale source masses introducing several technological advancements already available. Arguably, the most important change is the replacement of PMTs ---\thinspace which are the leading source of background in NEXT-100\thinspace--- with SiPMs, which are radiopure, pressure resistant and able to provide better light collection. Furthermore, it is possible to optimize the performance of the tracking signature through the operation of the detector with low-diffusion gas mixtures, resulting in better spatial resolution.\cite{NEXT-HD} We call \emph{high definition} (HD) this incremental approach to a tonne-scale detector. Monte-Carlo simulations show that the specific background rate of NEXT-100 could be reduced by at least one order of magnitude in a tonne-scale NEXT-HD, thus improving by more than one order of magnitude the current half-life limits.

A more radical approach to a tonne-scale experiment may be possible with the detection with high efficiency of the Ba$^{++}$ ion produced in a $^{136}$Xe \bbonu\ decay using \emph{single-molecule fluorescence imaging} (SMFI). The detection would occur in (delayed) coincidence with the identification of the two electrons and would ensure a background-free experiment. The possibility of using SMFI as the basis of molecular sensors for barium tagging was proposed in 2016,\cite{Jones:2016qiq} followed shortly after by a proof of concept which managed to resolve individual Ba$^{++}$ ions on a scanning surface using an SMFI-based sensor.\cite{McDonald:2017izm} Intense R\&D is under way in the NEXT Collaboration to implement a SMFI-sensor prototype capable of demonstrating barium detection. New molecular indicators able to provide an intense fluorescence signal in dry medium were presented recently.\cite{NEXT-BOLD} Although many steps need to be taken to demonstrate a full barium tagging sensor, the consistent success of the R\&D initiated in 2016 offers good prospects. We call this approach \emph{Barium On Light Detection} (BOLD). A NEXT-BOLD module would measure the energy and event position in the anode (with a SiPM array), reserving the cathode for the barium sensor. The delayed coincidence would permit relaxing the stringent topological restrictions imposed to the events in NEXT-100 (and NEXT-HD), resulting in a higher signal efficiency in addition to a negligible background index. A NEXT-BOLD module with a mass in the tonne range could reach a half-life sensitivity of about $10^{28}$~years.


\section*{References}
\begin{thebibliography}{99}

\bibitem{Weinberg:1979sa}
	S.~Weinberg,
	{\it Phys.\ Rev.\ Lett.} {\bf 43} (1979) 1566.

\bibitem{seesaw}
%
	P.~Minkowski,
	{\it Phys.\ Lett.} {\bf 67B} (1977) 421.

	M.~Gell-Mann, P.~Ramond and R.~Slansky,
	{\it Conf.\ Proc.\ C} {\bf 790927} (1979) 315.

	T.~Yanagida,
	{\it Conf.\ Proc.\ C} {\bf 7902131} (1979) 95.

	R.~N.~Mohapatra and G.~Senjanovic,
	{\it Phys.\ Rev.\ Lett.} {\bf 44} (1980) 912.

\bibitem{Fukugita:1986hr}
	M.~Fukugita and T.~Yanagida,
	{\it Phys.\ Lett.\ B} {\bf 174} (1986) 45--47.

\bibitem{GomezCadenas:2011it}
	J.J.~G\'omez-Cadenas, J.~Mart\'in-Albo, M.~Mezzetto, F.~Monrabal and M.~Sorel,
	{\it Riv.\ Nuovo Cim.} {\bf 35} (2012) 29.

\bibitem{KamLAND-Zen:2016pfg}
	A.~Gando {\it et al.} (KamLAND-Zen Collaboration),
	{\it Phys.\ Rev.\ Lett.} {\bf 117} (2016) 082503.
	Addendum: {\it Phys.\ Rev.\ Lett.} {\bf 117} (2016) 109903.

\bibitem{DBDM}
	V.~\'Alvarez {\it et al.} (NEXT Collaboration),
	{\it Nucl.\ Instrum.\ Meth.\ A} {\bf 708} (2013) 101.

	J.~Renner {\it et al.} (NEXT Collaboration),
	{\it Nucl.\ Instrum.\ Meth.\ A} {\bf 793} (2015) 62.

\bibitem{DEMO}
%
	V.~\'Alvarez {\it et al.} (NEXT Collaboration),
	{\it JINST} {\bf 8} (2013) P04002

	V.~\'Alvarez {\it et al.} (NEXT Collaboration),
	{\it JINST} {\bf 8} (2013) P05025.

	V.~\'Alvarez {\it et al.} (NEXT Collaboration),
	{\it JINST} {\bf 8} (2013) P09011.

	D.~Lorca {\it et al.} (NEXT Collaboration),
	{\it JINST} {\bf 9} (2014) P10007.

	L.~Serra {\it et al.} (NEXT Collaboration),
	{\it JINST} {\bf 10} (2015) P03025.

	P.~Ferrario {\it et al.} (NEXT Collaboration),
	{\it JHEP} {\bf 1601} (2016) 104.

\bibitem{Monrabal:2018xlr}
	F.~Monrabal {\it et al.} (NEXT Collaboration),
	{\it JINST} {\bf 13} (2018) P12010.

\bibitem{NEXT-White:2018}
	G.~Mart\'inez-Lema {\it et al.} (NEXT Collaboration),
	{\it JINST} {\bf 13} (2018) P10014.

	J.~Renner {\it et al.} (NEXT Collaboration),
	{\it JINST} {\bf 13} (2018) P10020.

	A.~Sim\'on {\it et al.} (NEXT Collaboration),
	{\it JINST} {\bf 13} (2018) P07013.
	
	P.~Novella {\it et al.} (NEXT Collaboration),
	{\it JHEP} {\bf 1810} (2018) 112.

\bibitem{NEXT-White:2019}
	J.~Renner {\it et al.} (NEXT Collaboration),
	arXiv:1905.13110 [physics.ins-det].

	P.~Ferrario {\it et al.} (NEXT Collaboration),
	arXiv:1905.13141 [physics.ins-det].

	P.~Novella {\it et al.} (NEXT Collaboration),
	arXiv:1905.13625 [physics.ins-det].

\bibitem{Martin-Albo:2015rhw}
	J.~Mart\'in-Albo {\it et al.} (NEXT Collaboration),
	{\it JHEP} {\bf 1605} (2016) 159.

\bibitem{NEXT-HD}
	C.~A.~O.~Henriques {\it et al.} (NEXT Collaboration),
	{\it Phys.\ Lett.\ B} {\bf 773} (2017) 663.

	J.~Renner {\it et al.} (NEXT Collaboration),
	{\it JINST} {\bf 12} (2017) T01004.

	R.~Felkai {\it et al.},
	{\it Nucl.\ Instrum.\ Meth.\ A} {\bf 905} (2018) 82.

	C.~A.~O.~Henriques {\it et al.} (NEXT Collaboration),
	{\it JHEP} {\bf 1} (2019) 027.
	
	A.~D.~McDonald {\it et al.} (NEXT Collaboration),
	{\it JINST} {\bf 14} (2019) P08009.

	A.M.F. Fernandes et al. (NEXT Collaboration), arXiv:1906.03984 [physics.ins-det]. 

\bibitem{Jones:2016qiq}
	B.~J.~P.~Jones, A.~D.~McDonald and D.~R.~Nygren,
	{\it JINST} {\bf 11} (2016) P12011.

\bibitem{McDonald:2017izm}
	A.~D.~McDonald {\it et al.} (NEXT Collaboration),
	{\it Phys.\ Rev.\ Lett.} {\bf 120} (2018) 132504.

\bibitem{NEXT-BOLD}
	P.~Thapa {\it et al.},
	arXiv:1904.05901 [physics.ins-det].
	
	I.~Rivilla {\it et al.},
	arXiv:1909.02782 [physics.ins-det].


\end{thebibliography}
\end{document}